\documentstyle[pre,aps]{revtex}

\begin{document} 
\twocolumn
\begin{center}
{\bf \Large Ultrametric space cut  instead of 2d one in 2d 
quantum field models}\\
\vspace{4mm}
D.B. Saakian\\

Yerevan Physics Institute,
Alikhanian Brothers St. 2,\\ Yerevan 375036, Armenia 

\end{center}

\begin{abstract}

It is possible to formulate 2d field theory on the ultrametric space 
with the free correlators identical to 2d correlators and the same 
potential. Such model should carry some features of original model for 
scale invariant theories. For the case of strings and 2d conformal
models it is possible to derive exact results. It is possible to
investigate not only  bulk structure (phase transition points) of theory,
but sometimes also correlators. Such ultrametric models could be 
naturally expressed via random energy model and directed polymer on 
Cayley tree.

\end{abstract}

REM [1-8] is one of fundamental models of modern physics. It was initially
introduced as the simplest model of spin glass, but later on its 
relation to a whole series of science fields were disclosed. This model 
has a decisive property of operation as an optimum coding device [2-4]. 
In [9-11] a relation of 2d quantum Liouville model to REM and to the 
Directed Polymer (DP) on Cayley tree. Using similar ideas we intend to 
construct 2d quantum models in ultrametric space, as well as bring out 
some results for strings.
Let us consider the ultrametric (UM) space with some measure for the 
 (sphere) area $d\mu_s(V,X)$, volume (ball) measure $d\mu_s(V,X)$,
total surface $e^V$ and total volume $e^V-1$. We can construct such UM 
space as a limit of hierarchic lattices. Consider a tree with a constant
number of branchings  q in each node  and number N for hierarchies. The 
number of end points is  $q^N$, the number of branches - 
$\frac{q^{N}-1}{q-1}$. We consider a set of end points as a surface of 
sphere, the set of branches makes a volume (of the ball), each point on 
the surface is connected with the origin  (zero level of the hierarchi) 
via a single path. We determine a measure $d\mu_s(V,X)=1$  for each end 
point, then $d\mu_l(V,l)=(q-1)$ for each link. Instead of integer
q  consider the limit $q\to 1, N\ln q\to V_0$. Now
we have for the total area $\mu_s=\int d\mu_s(V_0,X)=e^{V_0}$
 and for the total volume $\mu_l=\int d\mu_l(V,l)=e^{V_0}-1$. 
 We determine UM distance between two points x, y on this surface 
 $V_0-v\equiv V_0-\frac{n}{N}$, where n is the number of the hierarchic 
level, on which x and y had the last common node on trajectories to their
point from origin. The maximum UM distance between two points v on the
surface is $V_0$ (an ordinary distance as a function of V will be defined
lately). \\
We define some field $\phi(x)$ on our surface.
For determination of kinetic energy (quadratic form with the
Laplacian as the kernel in the conventional space) let us consider the 
expansion 
\begin{equation}
\label{e1}
\phi(X)=f_0+\int_0^{V_0}{\it d}Vf(V,l)
\end{equation}
Here $f(l)$ is determined on the links. The integration in (1) is made 
along the trajectory of point X. Since the measures on both the threads
of (1) coincide ($\int d\mu_l(V,l)=1+\int dV d\mu_l(V,l)$), we
omit the Jacobian.
Now determine the kinematic part of the action for the field $\phi(x)$ 
\begin{equation}
\label{e2}
\int_0^{V_0}{\it d}V{\it d}\mu_l(V,l)2\pi f(V,l)^2
\end{equation}
Then the partition under the potential UM 
\begin{equation}
\label{e3}
\begin{array}{l}
\int {\it d}f\exp\{-\int_0^{V_0}{\it d}V\int{\it d}\mu_l(V,l)2\pi
f(x,V)^2\}\\
\exp\{\int d\mu_s(V_0,X)U(\phi(X))\}
\end{array}
\end{equation}
We have for the correlator 
\begin{equation}
\label{e4}
<\phi(X)\phi(X')>=\frac{V}{4\pi}
\end{equation}
where V is the UM distance between the points X, X'. During the treatment
of (4) we have omitted the zero modes $f_0$ in (1) for $\phi(x)$. For 
usual 2d models with 
\begin{equation}
\label{e5}
\int {\it d}\phi_0{\it d}\phi\exp\{-\frac{1}{2} {\it d}x^2\nabla \phi(x)^2\}
\exp\{\int d x U(\phi(x))\}
\end{equation}
we have for the total surface $\pi R^2$ and for correlators 
\begin{equation}
\label{e6}
<\phi(X)\phi(X')>=\frac{\ln r}{2\pi}
\end{equation} 
We can determine the distance from the equality $V=\ln\pi r^2$. Then our
correlators coincide (in any case when $r\gg 1$). What is the advantage of
representation (3)? We are in a position to calculate the partition 
through the iterations. Let us take some large number K and determine the
iterations 
\begin{equation}
\label{e7}
\begin{array}{l}
I_1(x)=\sqrt{2K}\int _{-\infty}^{\infty}\exp\{-2\pi K y^2+U(x+y)\}{\it d }y\\
I_{i+1}(x)=\sqrt{2K}\int _{-\infty}^{\infty}\exp\{-2\pi K y^2\}
I_{i}^{\exp(\frac{V}{K})}(x+y){\it d }y\\
Z=\lim_{K\to \infty}I_{K}^{\exp(\frac{V}{K})}(0)
\end{array}
\end{equation}
As virtually for determination of partition we need only the equation (7),
our (speculation ) abstraction $q\to 1$ is quite correct. In
principle, in equation (2)
instead of dV one can take any measure  ${\it d}Vu(V)$, e.g.,
$u(V)=e^{\alpha V}$. In this case (the area considered by us $e^V\sim R^2$
the equation is transformed to
\begin{equation}
\label{e8}
<\phi(X)\phi(X')>\sim r^{-\frac{\alpha}{d}}
\end{equation}
So choosing $\alpha=d(d-2)$ we have d dimensional free field correlator. 
Now in (7) we replace $\exp(-2\pi K y^2)$ by
\begin{equation}
\label{e9} 
\begin{array}{l}
\exp(-2\pi K y^2u((i-1)/K))
\end{array}
\end{equation}
It is held that the equation of iteration is easily solved only when the
function $u(v)$ is constant (in our case (6)) or when u(v) steadily falls. 
In this case the SG physics corresponds to the pattern of gradual freezing
of the hierarchy levels [6]. In principle, one can repeat the same 
reasoning for fermions also. Let us introduce the fields  
$\psi(V,l),\bar\psi(V,l)$. Then  
\begin{equation}
\label{e10}
\begin{array}{l}
\int_0^{V_0}{\it d}V{\it d}\mu_l(V,l)\bar\psi(V,l)\psi(V,l)e^V\\
\bar\psi(V,X)\psi(V',X')+\psi(V,X)\bar\psi(V,X)=\delta(x-x')\delta(V-V')
\end{array}
\end{equation}
This will give an expression for correlators 
$\Psi(V_0,X)=\int_0^{V_0} d V\psi(V,l),
\bar\Psi(V_0,X)=\int_0^{V_0} d V\bar\psi(V,l)$
\begin{equation}
\label{e11}
<\bar\Psi(V_0,X)\Psi(X')>\sim r^{-1}
\end{equation}
To apply these ideas to the strings one is to remember that in [10] it 
was obtained how the correlator in directed polymer (DP) on Cayley tree 
is related to correlators in the Liouville model. It is seen that the 
strings are closely related to the UM space approach, REM and DP. For
 string's (corresponding to the 2d gravitation with matter field c) partition
it is known  that [12-15]
\begin{equation}
\label{e12}
\begin{array}{l}
Z=
\frac{1}{V_{SL(2,C)}}
\int_0^{\infty}{\it d}Ae^{-\mu A}Z(A)\\
Z(A)=\int D_g\phi\delta(\int d^2w \sqrt{\hat g}e^{\alpha \phi}-A)e^{\frac
{1}{8\pi}\int d^2w\sqrt{\hat g}{\phi \Delta \phi +QR\phi}}
\end{array}
\end{equation}
where $\frac{1}{V_{SL(2,C)}}$ is some factor connected with the fixing 
of gauging.  It is known for the coefficients 
\begin{equation}
\label{e13}
\begin{array}{l}
c=1-12\alpha_0^2\\
Q=2\sqrt{2+\alpha_0^2}\\
\alpha=-\frac{\sqrt{25-c}}{\sqrt{12}}+\frac{\sqrt{1-c}}{\sqrt{12}}\\
\frac{Q}{\alpha}=\frac{1}{12}[c-25-\sqrt{(25-c)(1-c)}]\\
\frac{1}{8\pi}\int d^2w\sqrt{\hat g}R=1
\end{array}
\end{equation}
The last equation is written for the sphere where the measure Dg is linear.
If we use the technique of zero mode, we can decompose $\phi(x)$ in two 
parts  $\phi\to \phi_0+\phi$, where $\phi_0$ is the zero mode. Here it 
is easy to make integration over  $\phi_0$ and obtain 
\begin{equation}
\label{e14}
Z=
\frac{1}{V_{SL(2,C)}}
\mu^{-\frac{Q}{\alpha}}\Gamma(\frac{Q}{\alpha})
\int D_g\phi e^{\frac
{1}{8\pi}\int d^2w\sqrt{\hat g}{\phi \Delta \phi +QR\phi}}(\int d^2w 
\sqrt{\hat g}e^{\alpha\phi})^{-\frac{Q}{\alpha}}
\end{equation}
We see that  $Z=<z^\mu>, \mu=-\frac{Q}{\alpha}$ and 
$z=\int d^2w\sqrt{\hat g}e^{\alpha\phi}$ resembles partition in some sense.
On the language of spin glasses this expression corresponds to real 
positive replica (for $0<c<1 \qquad \mu $ changes in interval $ 2<\mu<2.5$ )
In case of other topologies with $h>0$ we have to take
$\mu=-\frac{Q(1-h)}{\alpha}$.\\
Let us return to UM space in (14). Here instead of $d^2w\sqrt{\hat g}$ 
the UM measure $d \mu(X,V_0)$ is taken, and instead of  
$\frac{1}{8\pi}\int d^2w\sqrt{\hat g}\phi \Delta \phi $ quadratic form 
(3). Then one can strictly prove in analogy to [7] that the thermodynamics 
(14) (in the UM version) is equivalent to REM for final replicas [8]. 
Dividing 2d space as a whole in M parts, calculating the distribution (11),
and neglecting in the thermodynamic limit the term in the action Q term
(this term may turn out to be crucial at the determination of correlators)
we have
\begin{equation}
\label{e15}
\begin{array}{l}
<\delta(E-\phi(x)>=e^{-\frac{E^2}{2G(0)}}\\
G(0)=\ln {M}
\end{array}
\end{equation}
where G(r) is a correlator of the scalar field. Instead of our system in 
UM space we take a common REM (the case of final replicas) with M 
configurations and distribution (15) for $E_i$. Derrida and Gardner 
solved such a system [8]. Here we give only a qualitative derivation of 
exact results. 
\begin{equation}
\label{e16}
<(\sum_{i=1}^Me^{-\beta E_i})^{\mu}>
\end{equation}
(16) is calculated for positive integral values of $\mu$, 
where the averaging is made over the distribution (15) for each $E_i$. 
There are only two competing terms,- the first one has a form of 
$e^{-\beta{\mu} E_i}$  
\begin{equation}
\label{e17}
\begin{array}{l}
Z=<(\sum_{i=1}^Me^{-{\mu}\beta E_i})>=Me^{\frac{G(0)(\beta\mu)^2}{2}}\\
\ln Z=\ln M+\frac{G(0)\beta^2\mu^2}{2}
\end{array}
\end{equation}
and is one of diagonal terms in (16), and the second one is from the 
cross terms in expansion Z 
\begin{equation}
\label{e18}
\begin{array}{l}
Z=(<(\sum_{i=1}^Me^{-\beta E_i})>)^{\mu}=
M^{\mu}e^{\mu\frac{G(0)(\beta)^2}{2}}\\
\ln Z=\mu\ln M+\frac{G(0)\beta^2\mu}{2}
\end{array}
\end{equation}
 We consider these expressions in the thermodynamic 
limit. When the 
entropy $\ln Z-\frac{dZ}{d\beta}\beta$ in (17) 
is zeroth, then below the zeroing temperature ($T=\frac{1}{\beta}$) we take 
for $\ln Z$ the expression that is linear in $\beta$ and is continuous in the transition point. Eventually 
we have for the third phase
\begin{equation}
\label{e19}
\begin{array}{l}
\ln Z={\mu\sqrt{\beta_c}\beta}\\
\beta_c=\sqrt{\frac{2\ln M}{G(0)}}
\end{array}
\end{equation}
The third phase takes place for $\mu<1$. For $0<d<1$ and so  $2<\mu<2.5$ there occur
only the  phases (17)-(18).
So, for spherical topology our string is always in phase (17) (even at $d>1$ 
when our formulas (13) give complex values for $\beta$). \\
For  other topologies the string is in the phase (18) for $d<1$.
\\
Let us analytically continue (13) for $c>1$. Here both of our
parameters $\mu,\beta$ are complex. For higher
dimensions the string is first is in SG phase. As is definitely known from 
 SG physics, there should be some probabilistic description of a physical state state. 
When we continue increase the dimensionality, then at $d=19$ there is another 
phase transition. It is known from REM, that at complex temperatures 
$\beta=\beta_1+i\beta_2$ there is some new, Lee-Yang-Fisher phase, when  
\begin{equation}
\label{e20}
\begin{array}{l}
\beta_1=\frac{\beta_c}{2}\\
\beta_2>\frac{\beta_c}{2}
\end{array}
\end{equation}
This expression could be derived by means of representation 
$<\ln |Z|>=<\frac{|Z|^{\mu}-1}{\mu}>$. In calculations of 
$<|Z|^{\mu}>$ in $\mu\to 0$ limit  at complex $\beta$ the principal terms in (16) are the
crossing ones 
$e^{
-\beta_1( E_{i_1}+..E_{
i_{\frac{\mu}{2}}
})
}$.\\
From expression (13) for $\alpha\equiv \beta_1+i\beta_2$ we see
 that this happens at $c=19$. So at $d=19$ there is a second phase 
transition for strings (after first one at $d=1$) and the
topology higher than spherical.
We can apply the same ideas to other conformal theories using the 
Coulomb-gas formalism with the  background charge $\alpha_0$ [16].
To calculate a correlator
$\prod_k\exp\{i\alpha_k\}$
with
screening charges $Q_+^mQ_-^n$ one has consider
\begin{equation}
\label{e21}
\begin{array}{l}
Z=\int D_g\phi e^{\frac
{1}{8\pi}\int d^2w\sqrt{\hat g}{\phi \Delta \phi
+i2\sqrt{2}\alpha_0R\phi}}\\
\prod_k\exp\{i\alpha_k\}
(\int d^2 w\exp(i\alpha_+)^n
(\int d^2 w\exp(i\alpha_+)^m
\end{array}
\end{equation}
We see in the  ultrametric space this is
equivalent to the system of 2 real replica REM-s at different imaginary 
temperatures.\\
When one considers the correlators without screening charges,
 the scaling indices of the  2 point correlator are the same as those in the 2d
 case. It is possible to derive minimal models also in this
 scheme. We can  calculate the multipoint correlators and compare
  with 2d ones,bearing in mind that  in our UM space all triangles are isosceles.
It is interesting to compare the structural couplings for 2d and ultrametric cases.
It will be interesting to apply this approach to other exactly solvable models
in 2d as well as to other versions of strings.\\
As for the construction of theories in dimensions $d>2$
n principle there is a phenomenological criteria for establishment of relation of any
 theory to REM and eventually to our ultrametric space approach.
It is that, if the free energy of system is a logarithm 
of the  number of particles. In REM the efficient number of couplings is 
an exponent of the number of spins or free energy. The same situation is
with scale invariant theories in any dimensions. Another argument is that in critical 
theories system can  ignore some details of geometry and so, do not 
distinguish real space from our UM space.
So there is some chance
that at the critical point our ultrametric space approach could work also at 
$d>2$. This hypothesis should be checked by means of Monte-Carlo simulations.\\
The ultrametric space approach is somewhat  crude (I don't see
holomorphic+antiholomprphic decomposition and Majorana fermions,
as in 2d space), but is a  complete concept. It could will be successsful, when the real
space picture is too complicated (strings, $d>2$ critical
theories, turbulence) and the main problem consists in
 the closure.\\
I am grateful to S. Apikyan, D. Karakhanyan,C. Lang, R. Mkrtchyan,
 V. Ohanyan, A.Sedrakyan
 for discussions, Amsterdam University for partial financial support.

\end{document}